\documentclass[prb,twocolumn,superscriptaddress, preprintnumbers,amsmath,amssymb]{revtex4}

\usepackage{graphicx,color}
\graphicspath{{./Figures/}}
\usepackage{multirow}
\usepackage{braket}

\begin{document}

\title{Carbon nanotube quantum dots on hexagonal boron nitride}

\author{A. Baumgartner}
\affiliation{
Institute of Physics, University of Basel, Klingelbergstrasse 82, 4056 Basel, Switzerland\\}
\email{andreas.baumgartner@unibas.ch}
\author{G. Abulizi}
\affiliation{
Institute of Physics, University of Basel, Klingelbergstrasse 82, 4056 Basel, Switzerland\\}
\author{K. Watanabe}
\affiliation{
National Institute for Material Science, 1-1 Namiki, Tsukuba, 305-0044, Japan\\}
\author{T. Taniguchi}
\affiliation{
National Institute for Material Science, 1-1 Namiki, Tsukuba, 305-0044, Japan\\}
\author{J. Gramich}
\affiliation{
Institute of Physics, University of Basel, Klingelbergstrasse 82, 4056 Basel, Switzerland\\}
\author{C. Sch\"onenberger}
\affiliation{
Institute of Physics, University of Basel, Klingelbergstrasse 82, 4056 Basel, Switzerland\\}

\begin{abstract}

We report the fabrication details and low-temperature characteristics of the first carbon nanotube (CNT) quantum dots on flakes of hexagonal boron nitride (hBN) as substrate. We demonstrate that CNTs can be grown on hBN by standard chemical vapor deposition and that standard scanning electron microscopy imaging and  lithography can be employed to fabricate nanoelectronic structures when using optimized parameters. This proof of concept paves the way to more complex devices on hBN, with more predictable and reproducible characteristics and electronic stability.
\end{abstract}

\maketitle


Carbon nanotubes (CNTs) are a versatile fundamental building block for classical small scale electronics and quantum electronics,\cite{Laird_Kouwenhoven_arXiv_2014} and for the investigation of novel quantum states.\cite{Schindele_Baumgartner_PRB89_2014} However, the ideal properties of CNTs are usually masked by electrical potential fluctuations induced by the substrate.
An intuitive solution to avoid these fluctuations is to suspend the active device volume above the substrate, which has led to fundamental experiments, both on suspended CNTs\cite{Cao_Small_2005, Kuemmeth_Ilani_McEuen_Nature_2008, Waissman_Ilani_NatureNano_2013, Jung_Baumgartner_Nanolett_2013} and graphene.\cite{Rickhaus_Maurand_NatureComm_2013}

Suspended devices, however, suffer from limitations in the scalability, geometry and in the choice of the contact and gate materials. For example, it is difficult to find a superconductor or a ferromagnet suitable for the growth of CNTs by chemical vapor deposition (CVD) at temperatures around $1000^{\circ}$C. Stamping techniques\cite{Waissman_Ilani_NatureNano_2013, Kontos_APL} are more versatile, but depend strongly on the interface characteristics of the contacts. In contrast, devices on a substrate offer a much larger variety of design options and suitable materials, but the stability and quality of the electronic structure are usually compromised. Standard cleaning techniques, e.g. dry etching, cannot be deployed because they also remove the carbon structures,\cite{Samm_Gramich_Baumgartner_JAP_2014} while the thermal coupling to the substrate is too large for in-situ current annealing. 

For graphene, a new approach has recently led to 'clean' nanostructures, namely the use of thin layers (flakes) of hexagonal boron nitride (hBN) as insulating substrates. This approach allows the implementation of substrate supported graphene in high-mobility transistors\cite{Dean_Kim_NatureNano_2010} and the observation of fundamental physical phenomena.\cite{Yankowitz_LeRoy_NaturePhys_2012, Ponomarenko_Geim_Nature_2013} Using hBN as a substrate has also enabled the fabrication of highly efficient monolayer WSe$_2$ light emitting diodes.\cite{Ross_NatureNano_2014} In contrast to the standard Si/SiO$_2$ substrates, hBN exhibits significantly less charge traps and leads to larger charge puddles in graphene\cite{Xue_LeRoy_NatureMat10_2011} and a reduction of electronic instabilities, e.g. in graphene quantum dots (QDs).\cite{Engels_Stampfer_APL103_2013} hBN can be cleaved (mechanically exfoliated) by simple methods,\cite{Dean_Kim_NatureNano_2010} with resulting thicknesses down to single atomic layers. To date, big efforts are made to fabricate hBN-graphene multi-layer structures,\cite{Britnell_Eaves_NatureComm4_2012} one dimensional contacts to graphene\cite{Wang_Dean_Science342_2013} and combinations with other layered materials.

\begin{figure}[b]
	\centering
		\includegraphics{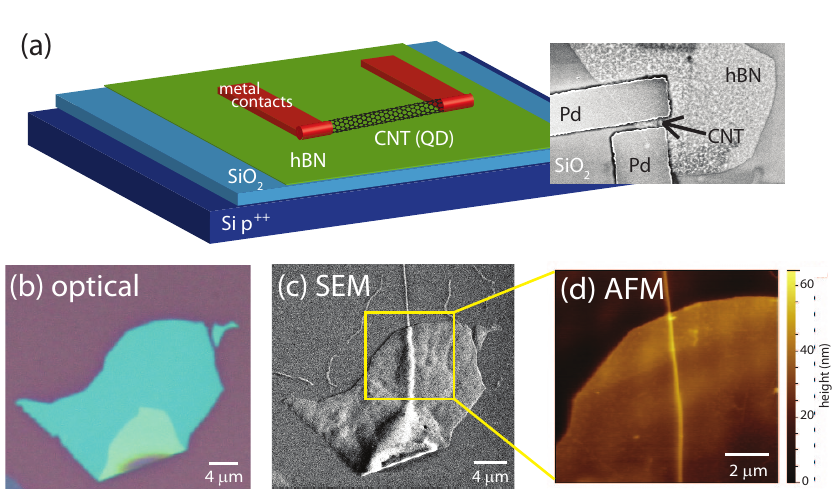}
	\caption{(Color online) (a) Schematic of a CNT QD structure on an hBN flake. The inset shows an SEM image of a device. (b) optical, (c) SEM and (d) AFM image of a CNT ($\sim 8\,$nm radius) on an hBN flake ($\sim 28\,$nm thickness) on a SiO$_2$ substrate.}
\end{figure}

So far, the use of hBN as substrate for more complex nanostructures has not been demonstrated. Specifically, no experiments on CNTs grown on hBN substrates have been reported. The main reason is probably that CNTs are difficult to locate on hBN, because optical microscopy lacks the required resolution, scanning electron microscopy (SEM) images can be of poor quality (see below), and imaging by atomic force microscopy (AFM) is rather demanding because of the large lateral and vertical scales involved, while requiring nanometer resolution to image CNTs. Here we report the fabrication details and low-temperature characteristics of a CNT QD on top of an hBN flake. We demonstrate that for a range of hBN thicknesses and SEM settings, rapid feed-back and large scale SEM imaging of CNTs on hBN are possible, also shedding light on the contrast mechanisms when imaging nano-objects on dielectrics. Based on these results, we fabricate CNT QDs on hBN and report first low-temperature characteristics, i.e. the formation of a 'clean' QD. This fabrication technique on hBN is easily applicable to more complex devices, similar as on standard substrates, which suggests that hBN can easily be used as substrate for a variety of other nanostructures.

The structure of our devices is shown schematically in Fig.~1a. We use a highly p-doped Si wafer with a thermally oxidized $300\,$nm thick insulation layer, which allows one to use the substrate as a backgate. We deposit hBN flakes by mechanical exfoliation from a single crystal and achieve a moderate control over the thicknesses by adjusting the number of sequential exfoliation steps. We obtain a rough thickness estimate and fast feed-back using an optical microscope.\cite{Gorbachev_Geim_Small_2011, Golla_LeRoy_APL102_2013} Ru/Fe catalyst particles\cite{Li_Dai_JACS_2007} are deposited on the wafer surface and on the hBN flakes by spin coating. The CNTs are grown at 850$^{\circ}$C in a CVD process with Methane as the source gas. Subsequently, $80\,$nm thick Pd contacts are fabricated by standard electron beam lithography using an optimized recipe for residue-free polymer removal.\cite{Samm_Gramich_Baumgartner_JAP_2014}

\begin{figure}[t]
	\centering
		\includegraphics{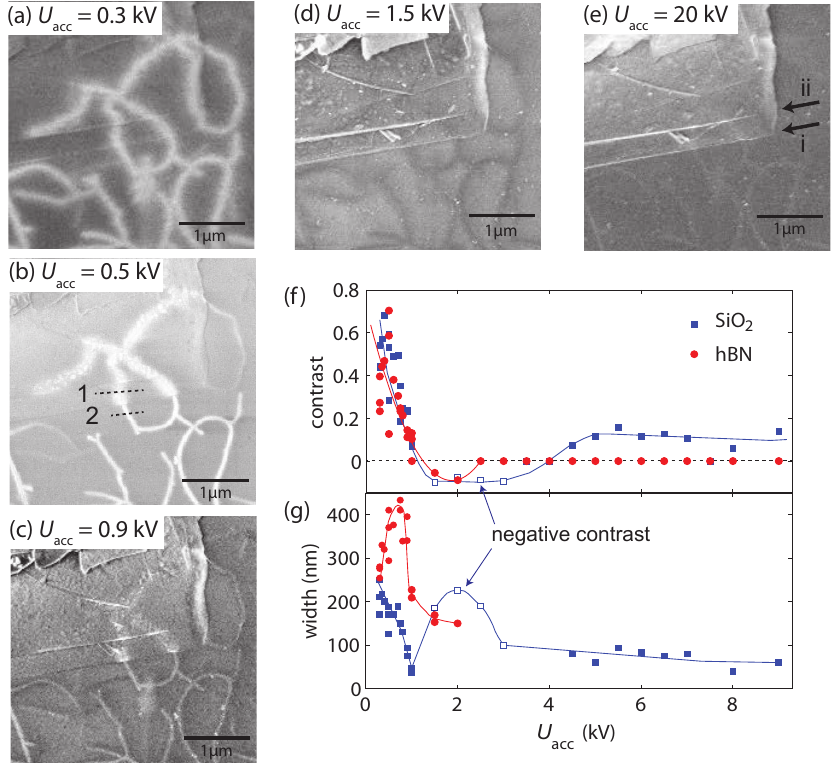}
	\caption{(Color online) (a)-(e) SEM images of CNTs on an hBN flake for different SEM acceleration voltages $U_{\rm acc}$. The thickness steps i and ii indicated in (e) are $\sim 35\,$nm and $\sim 80\,$nm, respectively. (f) SEM contrast and (g) apparent CNT width as a function of $U_{\rm acc}$ found in cross sections 1 and 2 indicated in (b). The continuous lines are guides to the eye.}
\end{figure}

Figure~1b shows the optical microscopy image of a $\sim28\,$nm thick hBN flake after CNT growth. The contrast and color allow for an initial screening for suitable flakes on a marker field before the device fabrication. The CNTs are not visible using an optical microscope and can only be found by SEM or AFM imaging, as demonstrated for the same hBN flake in Figs.~1c and 1d, respectively. The images show a CNT with a radius of $\sim 8\,$nm on the $\sim 28\,$nm thick hBN flake. The CNT radius and hBN thickness we determine from corresponding AFM images. Since AFM imaging is slow and restricted to rather small scan ranges, we have optimized the SEM parameters with the goal of obtaining simultaneous image contrast for both, hBN flakes and CNTs. For flakes thinner than $\sim40\,$nm we often find CNTs on hBN flakes suitable for device fabrication. The SEM contrast of hBN and CNTs depend crucially on the SEM electron acceleration voltage $U_{\rm acc}$. Figures~2a-2e show a series of SEM images at different $U_{\rm acc}$ of a $\sim 1\,$nm radius CNT lying partly on SiO$_2$ and partly on hBN (We use an in-lens detector, an aperture of 30$\,\mu$m and a primary electron beam current of $\sim1\,$nA). The hBN thickness in this image increases in two steps, first to $\sim 35\,$nm (arrow i in Fig.~2e) and then to $\sim80\,$nm (arrow ii).

For the lowest acceleration voltage shown in Fig.~2a, the hBN flake is barely visible, while the CNTs have the largest contrast of all investigated voltages (the flake position can be found by comparing to the other sub-figures). With increasing $U_{\rm acc}$ the flake becomes continuously easier to discriminate. For low $U_{\rm acc}$ the hBN bulk contrast is small and the flakes are visible mainly at the edges, consistent with a topographically determined emission of secondary electrons. The SEM contrast of the CNTs is more complex. For $U_{\rm acc}$ up to $\sim2\,$kV the contrast is similar for CNTs on hBN and directly on SiO$_{2}$. It is positive up to around $U_{\rm acc}=1\,$kV and negative at higher voltages. On hBN the contrast vanishes at $U_{\rm acc}\approx2.2\,$kV, while on SiO$_2$ it becomes positive again for $U_{\rm acc}>4\,$kV and remains roughly constant up to $U_{\rm acc}=20\,$kV, the maximum investigated voltage. The CNT contrast $(I_{\rm CNT}-I_{\rm sub})/(I_{\rm CNT}+I_{\rm sub})$  is plotted in Fig.~2f as a function of $U_{\rm acc}$ for the cross sections indicated in Fig.~2b, with the maximum intensities from the CNT and the substrate, $I_{\rm CNT}$ and $I_{\rm sub}$, respectively. 

The apparent CNT widths from the same image cross sections are plotted in Fig.~2g. The width is similar on both materials at the lowest voltages, but about four times larger on hBN than on SiO$_2$ around $U_{\rm acc}=0.7\,$kV. It also depends qualitatively different on $U_{\rm acc}$: on hBN it increases with $U_{\rm acc}$ at low voltages and then decreases at higher values. On SiO$_2$ the width continuously decreases with $U_{\rm acc}$ and becomes roughly constant at higher voltages. When the contrast is negative, the width changes differently. From these measurements we find an optimal $U_{\rm acc}$ between $0.6\,$kV and $1.0\,$kV for simultaneously imaging hBN flakes and CNTs. 

\begin{figure}[b]
	\centering
		\includegraphics{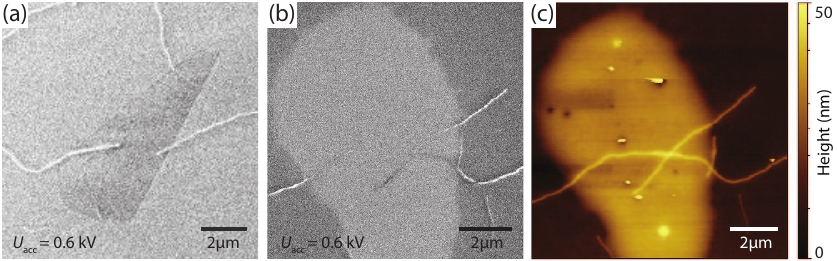}
	\caption{(Color online) SEM images of CNTs on hBN. (a) right tube: $\sim 5.5\,$nm radius, flake thickness $\sim 6\,$nm. (b) an $\sim8\,$nm radius CNT spanning a $\sim30\,$nm thick hBN flake. (c) AFM image corresponding to (b).}
\end{figure}

On flakes thinner than $\sim10\,$nm the apparent CNT diameter and contrast is almost identical on hBN and on the bare SiO$_2$. This is illustrated in Fig.~3a, which shows an SEM image of several CNTs on a $6\,$nm thick hBN flake. The CNT on the right side of the flake has a radius of $\sim5.5$nm. Though thin flakes lead to a better SEM contrast, the detrimental effects of the SiO$_2$ below the hBN will have an increased impact on an actual device.
That the contrast changes with the hBN thickness can be directly seen in Fig.~3b, with the corresponding AFM image in Fig.~3c. Two CNTs cross on top of the hBN flake, but both are visible in the SEM image only at the edges of the flake (bulk thickness $\sim30\,$nm) and on the SiO$_2$. The CNT spanning the whole flake has a radius of $\sim8\,$nm. At the edges of the flake, the hBN thickness increases continuously while the SEM contrast of the CNT is continuously diminished. We note that the first hBN step in Fig.~2 is also roughly $30\,$nm, but the SEM still shows a clear contrast for the CNT, suggesting a dependence of the contrast on the CNT diameter (CNT radius in Fig.~2 is $\sim1\,$nm, in Figs.~3b $\sim8\,$nm).
Generally, it is easiest to find CNTs that completely span a given hBN flake. These tubes probably grow vertically and then fall across the flake. However, we regularly find CNTs starting and ending on larger hBN flakes, suggesting that CNTs also grow directly on top of the flakes.


The contrast mechanism for SEM imaging of CNTs on insulating substrates\cite{Homma_APL_2004, Zhang_Nanotechnology17_2006} can be understood qualitatively in a simple picture: in the bare substrate the incident primary electrons (PEs) generate a large number of secondary electrons (SEs) in the dielectric at energies lower than the PEs, but larger than the material's energy gap. These SEs can leave the substrate through the surface or are absorbed in the material. The total charge of the layers depends on the balance between the number of injected PEs and emitted SEs. At low acceleration voltages, the PEs do not penetrate deep into the substrate and more SEs are emitted than absorbed, which leads to a positively charged surface layer.\cite{Glavatskikh_JAP89_2001} At higher voltages, the electrons penetrate deeper and leave the dielectric with reduced probability, which leads to a negative charging by the PEs. When the PE and SE penetration depths reach the insulator thickness, the SEs can be absorbed by the conducting backgate and the dielectric can again become positively charged.

Intuitively, the generation of SEs depends on the local electron density. The CNTs can be seen as charge reservoirs (or capacitors if not connected to an electrical contact) that supply electrons or accept electrons from the substrate, leading to an electric field determined by the surface charging and thus by the SEM acceleration voltage. Since the surface is insulating, the only mobile carriers are essentially the electrons excited to the conduction band by the SEM beam, which leads to the so-called electron beam induced conductance (EBIC), well known from semiconductor device characterization. These carriers spread from the CNT due to the electric field, which depends on the material's dielectric constant (Mott-Gurney law), until they thermalize and localize in the dielectric. For positively charged substrates, this leads to a relative increase in the local electron density and an increase in the SE generation rate, while the opposite happens when the substrate is negatively charged. The difference in the apparent CNT widths on SiO$_2$ and hBN can now be understood qualitatively by noting that hBN has an anisotropic dielectric constant: the component perpendicular to the substrate plane is $\epsilon_{perp}\approx5$, similar to SiO$_2$ ($\epsilon\approx4$), while the parallel component is $\epsilon_{par}\approx7$, leading to an increased EBIC parallel to the surface compared to SiO$_2$. We note that for suspended CNTs or at higher acceleration voltages, other mechanisms might come into play, for example the plasmon mediated generation of SEs.\cite{Kasumov_APL89_2006}

\begin{figure}[b]
	\centering
		\includegraphics{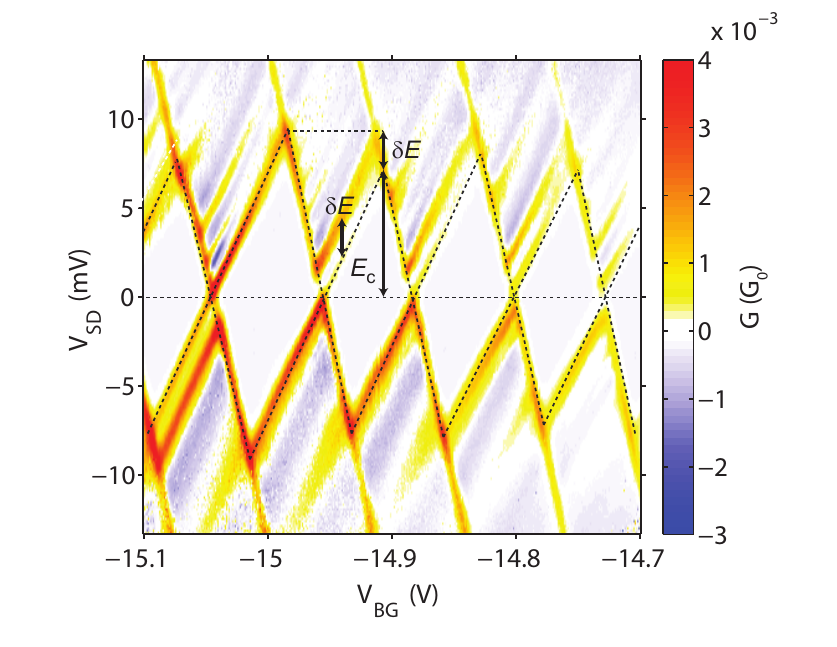}
	\caption{(Color online) Coulomb blockade and excited states resonances on a CNT QD fabricated on top an hBN flake ($T=4.2\,$K).}
\end{figure}

The fast and reliable SEM imaging of CNTs on hBN flakes allows the fabrication of nanoelectronic devices by standard electron beam lithography. We have used a recipe optimized to obtain polymer free CNTs and reliable CNT contacts\cite{Samm_Gramich_Baumgartner_JAP_2014} and thermal evaporation of Pd to fabricate two-terminal devices on CNTs on top of hBN flakes. In the device discussed here, the contact separation is $L\approx400\,$nm on a CNT of $r\approx5.5\,$nm radius on an hBN flake of $\sim6\,$nm thickness. The differential conductance of the device at $4.2\,$K (Helium bath) is plotted in Fig.~4 as a function of the backgate voltage $V_{BG}$ and the source-drain bias $V_{SD}$. Between the metal contacts a QD forms which leads to clear Coulomb blockade diamonds and a series of resonances due to excited states. The dashed lines in the figure trace the edges of the Coulomb blockade diamonds and suggest a four-fold symmetry, as expected for clean CNT QDs due to the spin and valley degeneracies in CNTs. The charging energy is $E_{\rm c}\approx7.2\,$meV, as indicated in Fig.~4, with a lever arm $\alpha_{\rm BG}$ similar to devices on SiO$_2$. From $E_{\rm c}=e^2/C_{\rm tot}$ we estimate the backgate capacitance $C_{\rm BG}\approx \alpha_{\rm BG} C_{\rm tot}$, in reasonable agreement with finite element method (FEM) calculations for a metallic cylinder with a length given by the contact distance.\cite{footnote1} This suggests that the QD confinement is determined by the metal contacts and not by defects in the CNT.

In Fig.~4 we observe up to the fifth excited state. The excited state energies $\delta E$ are roughly equidistant and similar for all Coulomb diamonds. We find $\delta E\sim2.2\,$meV either by the difference in the addition energies of the individual Coulomb diamonds, or by direct spectroscopy, as indicated in Fig.~4. Assuming a hard-wall confinement potential and a strong lifting of the sublattice band energies,\cite{Nygard_APA_1999} the level spacing is given by $\delta E=\hbar v_{\rm F}\cdot \pi/(2L)$, with $v_{F}\approx 8.1\times10^5\,$m/s the Fermi velocity. This reproduces the experiment for $L\approx380\,$nm, in good agreement with the contact spacing.

We note that one finds finite-bias regions of negative differential conductance and that the ground state transitions are weaker in some Coulomb blockade diamonds than the excited state transitions. Both findings suggest that the tunnel coupling of the excited states to the leads can be stronger than of the ground state, which leads to a competition for the QD occupation by the individual transmission channels. The fact that such detailed excited state spectroscopy is possible also supports the claim that the QD are 'clean' in the sense that no other electronic structures and resonances interfere with the ideal patterns. Specifically, the electron injection into the QD states is not mixed at the contact interfaces, which leads to the observed spectroscopy results. In addition, we reproducibly find a very good long term ($>$days) electrical stability, i.e. very few gate-dependent and no temporal charge rearrangements, comparable only with the best of our CNT QDs fabricated on SiO$_2$ substrates.

In summary we report detailed scanning electron microscopy imaging of carbon nanotubes on hexagonal boron nitride that allows to locate CNTs on hBN flakes. This is a fundamental prerequisit for a fast and reliable fabrication of standard top-down nanostructures, e.g. by electron beam lithography. We demonstrate a first CNT quantum dot on hBN, for which we discuss the electronic structure that indicates a very good electrical device quality and stability.

This work was supported by the EU FP7 Project SE2ND, the EU ERC
Project QUEST, the Swiss National Science Foundation (SNF), including
the projects NCCR QSIT and the NCCR Nano.

\bibliographystyle{apsrev}

\begin{thebibliography}{99}
\bibitem{footnote1}
\bibinfo{note}{For CNT QDs the source and drain capacitances can usually not be neglected and the often used formulas for a cylinder capacitor cannot be applied directly.}
\end{thebibliography}

\end{document}